\documentclass[journal]{IEEEtran}
\usepackage{amsmath,amsfonts}
\usepackage{algorithmic}
\usepackage{array}
\usepackage{textcomp}
\usepackage{stfloats}
\usepackage{url}
\usepackage{verbatim}
\usepackage{graphicx}
\usepackage{booktabs}
\usepackage{multirow}
\usepackage{bbding}
\usepackage[utf8]{inputenc}
\usepackage{mdframed}
\usepackage[ruled,boxed]{algorithm2e}
\usepackage{color}
\definecolor{bottomcolor}{RGB}{247, 248, 236}
\def\BibTeX{{\rm B\kern-.05em{\sc i\kern-.025em b}\kern-.08em
    T\kern-.1667em\lower.7ex\hbox{E}\kern-.125emX}}

\newcommand{\etal}{\textit{et al.}}

\begin{document}

\title{Generative AI Enables EEG Super-Resolution  \\ via Spatio-Temporal Adaptive Diffusion Learning}
\author{Shuqiang Wang, Tong Zhou, Yanyan Shen, Ye Li, Guoheng Huang, Yong Hu,
\thanks{Corresponding author: Ye Li, Email: ye.li@siat.ac.cn}
\thanks{Shuqiang Wang and Tong Zhou contributed equally to this work.}
\thanks{Shuqiang Wang, Tong Zhou, Yanyan Shen and Ye Li are with the Shenzhen Institutes of Advanced Technology, Chinese Academy of Sciences, Shenzhen 518055, China, and also with the University of Chinese Academy of Sciences, Beijing 100049, China.}
\thanks{Yong Hu is with Department of Orthopaedics and Traumatology, The University of Hong Kong, Hong Kong, China}
\thanks{Guoheng Huang is with the School of Computer Science and Technology, Guangdong University of Technology, Guangzhou, China}
}

\maketitle

\begin{abstract}
Electroencephalogram (EEG) technology, particularly high-density EEG (HD EEG) devices, are widely used in fields such as neuroscience. HD EEG devices improve the spatial resolution of EEG by placing more electrodes on the scalp, which meet the requirements of clinical diagnostic applications such as epilepsy focus localization. However, this technique faces challenges, such as high acquisition costs and limited usage scenarios. In this paper, spatio-temporal adaptive diffusion models (STAD) are proposed to pioneer the use of diffusion models for achieving spatial SR reconstruction from low-resolution (LR, 64 channels or fewer) EEG to high-resolution (HR, 256 channels) EEG. Specifically, a spatio-temporal condition module is designed to extract the spatio-temporal features of LR EEG, which then used as conditional inputs to direct the reverse denoising process.
Additionally, a multi-scale Transformer denoising module is constructed to leverage multi-scale convolution blocks and cross-attention-based diffusion Transformer blocks for conditional guidance to generate subject-adaptive SR EEG.
Experimental results demonstrate that the STAD significantly enhances the spatial resolution of LR EEG and quantitatively outperforms existing methods. Furthermore, STAD demonstrate their value by applying synthetic SR EEG to classification and source localization tasks, indicating their potential to Substantially boost the spatial resolution of EEG.
\end{abstract}


\section{Introduction}
\label{sec:introduction}
\begin{figure}[ht]
    \includegraphics[width=\linewidth]{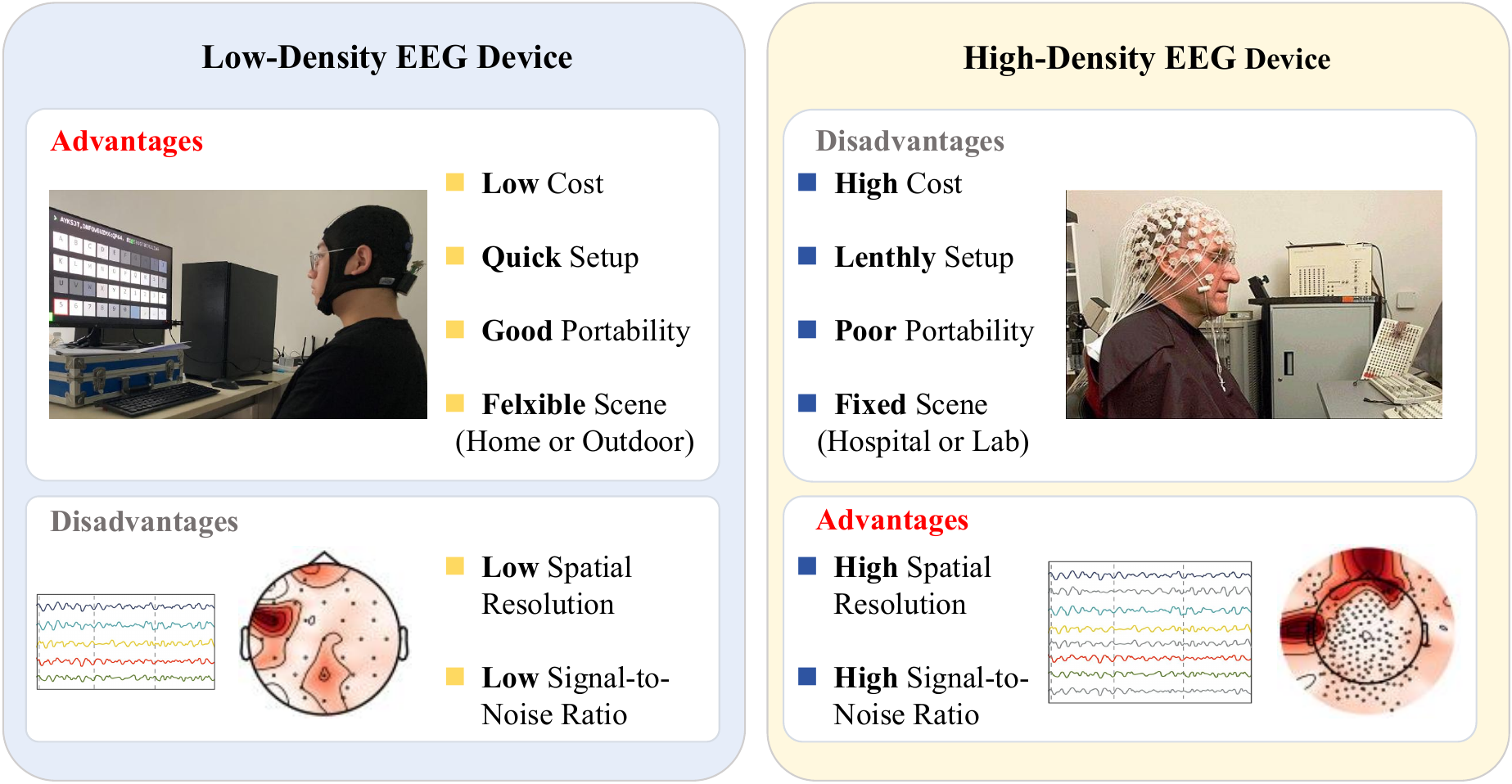}
    \caption{Comparative illustration of low-density and high-density EEG Devices: advantages and disadvantages. The picture on the right shows the buddhist monk Barry Kerzin meditating with a  high-density EEG device, picture from \cite{hanrath2019finite}.}
    \label{fig:intro}
\end{figure}

\IEEEPARstart{E}{lectroencephalogram} (EEG) is a economical, non-invasive neuroimaging technology with high temporal resolution, playing a vital role in cognitive neuroscience research and clinical diagnosis. Unlike other neuroimaging techniques such as positron emission tomography (PET) and functional magnetic resonance imaging (fMRI), EEG offers superior temporal resolution and can continuously monitor brain activity in natural environments. This capability substantially supports the study of dynamic cognitive processes \cite{eeg}. Recently, the development of portable and wireless EEG devices has further improved the flexibility and usability of EEG, enabling its widespread application in emerging neural engineering scenarios such as epileptic seizure detection and brain-computer interface (BCI) control \cite{portable1, portable2,xu2024neural}. Besides, these consumer devices can realize real-time remote monitoring of patients through the Internet of Medical Things (IoMT), enhance the early diagnosis and treatment, and improve the performance of real-time applications in intelligent consumer healthcare systems \cite{eseiz,TCE-chen}. However, the devices typically have fewer than 64 channels, resulting in lower spatial resolution of the acquired EEG data, which severely limits its potential in clinical diagnosis and brain function imaging research \cite{ldeeg_bad}. Currently, high-density (HD) EEG devices improve the spatial resolution of EEG by employing more electrode quantities, thus assisting doctors in performing evaluations or diagnoses in different clinical scenarios. For example, recent studies have shown that non-invasive HD EEG devices can objectively, non-invasively, and accurately determine the epileptic focus area \cite{hdeeg1, hdeeg2, hdeeg3}. However, HD EEG devices greatly increase hardware and acquisition costs, their bulkiness and complex wiring prevent subjects from moving their heads freely, restricting their use to controlled environments such as laboratories or hospitals. In addition, due to poor wearing comfort, subjects cannot be monitored for extended periods \cite{hdeeg_bad}. Fig. \ref{fig:intro} summarizes the respective advantages and limitations of different density EEG devices and reveals difficulties in acquisition and limited application scenarios for HD EEG devices.

The aforementioned observation highlights a significant dilemma in the current clinical diagnosis and treatment using EEG: due to high costs, discomfort from wearing, and other factors, high-resolution (HR) EEG is scarce. Consequently, low-resolution (LR) EEG is inadequate for meeting the demands of emerging neurotechnological applications and clinical requirements, such as the pre-surgery assessment of patients \cite{zhao2022tailoring}. In such cases, the EEG spatial super-resolution (SR) method offers an effective remedy. Generative AI is able to enhance the spatial resolution of the existing low-density EEG devices by transforming LR EEG into HR EEG, enabling improved brain analysis for applications like BCI, source localization, and sleep monitoring. For instance, EEG SR can reconstruct 256-channel EEG from 32-channel counterparts. Notably, EEG SR methods hold substantial potential for clinical applications, providing a cost-effective and efficient method to acquire HR EEG data.

Currently, in the field of enhancing spatial resolution of EEG, the methodologies predominantly fall into two main categories: channel interpolation based on mathematical models and EEG SR reconstruction utilizing deep learning methods. Channel interpolation methods principally rely on predefined scalp surface models to reconstruct missing channel data using signals from other channels, thereby improving spatial resolution. These methods involve calculating the second-order derivatives of the interpolation function in space to explore the distribution across the electrode channels \cite{Courellis_inter}. Subsequent studies have attempted more precise measurements of the spatial distances between EEG electrodes and more accurate predictions for missing channel data, for example, using inverse distance weighting based on Euclidean distances and multi-quadratic interpolation methods. However, these methods only utilize the spatial correlations among adjacent electrodes and typically struggle to effectively capture and reconstruct the high-frequency and transient components of brain electrical activity resulting in suboptimal reconstruction performance for non-stationary EEG signals.

Lately, deep learning has become the mainstream method in multiple fields, including the field of Internet healthcare \cite{TCE1,TCE2,TCE3} and medical image processing \cite{you2022fine, wang2020ensemble,TPAMI-zong}. Specifically, deep learning-based EEG SR technology involves training deep learning models to learn the implicit mappings between LR EEG and HR EEG, and using these trained models to reconstruct SR EEG. Notably, research has introduced EEG SR techniques based on convolution neural networks (CNN) \cite{yu2021morphological}, followed by studies proposing deep EEG SR frameworks using graph convolution networks (GCNs) \cite{zuo2021multimodal} to correlate the structural and functional connections of participants' brains \cite{tang_gcn}. Additionally, generative models offer a new learning paradigm for medical image reconstruction \cite{hu2021bidirectional}. Existing researches have utilized approaches such as Autoencoders (AE) and Generative Adversarial Networks (GANs) \cite{pan2024decgan} to enhance the spatial resolution of EEG. Although current EEG SR methods based on deep learning have made considerable advancements in enhancing EEG spatial resolution, they are still limited in effectively capturing both the temporal dynamics and spatial features of EEG, resulting in limited generalizability and practicality of the models. In addition, the enhanced spatial resolution achieved by these methods is limited, with a maximum capacity to reconstruct EEG data up to 64 channel level, which may not meet the requirements of specific clinical scenarios.

In clinical diagnostic and therapeutic settings, achieving high-fidelity reconstruction of HR (256 channels) ground truth EEG presents a substantial challenge. The key is accurately depicting the dependencies between LR EEG and HR EEG and addressing the significant channel-level disparity between them. Recently, a novel generative model—diffusion models \cite{ddpm} have exhibited remarkable generative capabilities in computer vision, leading to a significant increase in the Artificial Intelligence Generated Content (AIGC) \cite{aigc, gong2023generative}. Moreover, diffusion models have demonstrated superior image generation quality compared to GANs, while also providing advantageous features such as comprehensive distribution coverage, extensibility, and stable training targets \cite{ddpm_beat}.
Besides, Latent Diffusion Models (LDMs) \cite{rombach2022LDM} extend diffusion models into latent spaces, utilizing cross-attention mechanisms to achieve refined control in conditional generation tasks. /hl{Furthermore, SynDiff \cite{ozbey2023SynDiff} and DiffMDD \cite{wang2024diffmdd} have illustrated that diffusion models are applicable to generate medical images, thus advancing the precision of disease diagnosis. Despite this, applying diffusion models to enhance EEG spatial resolution has not been thoroughly explored. Therefore, this paper aims to investigate the potential of diffusion models in achieving EEG SR, particularly in enhancing spatial resolution. Through this novel generative model, we hope to provide a novel and effective method for processing and analyzing EEG data, supporting more accurate EEG studies and applications in various neuroscience contexts.}

\begin{figure*}[ht]
    \includegraphics[width=\linewidth]{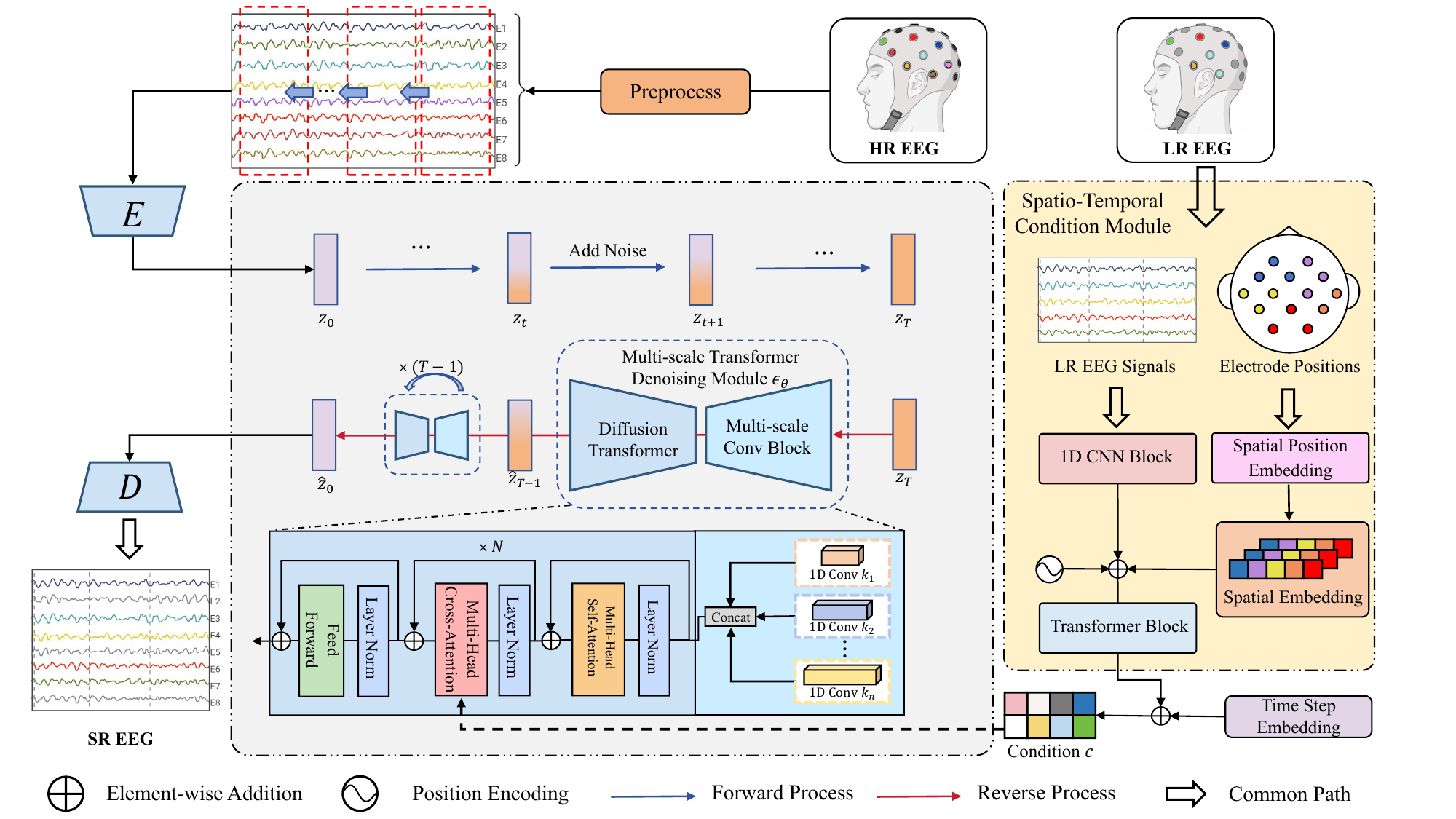}
    \caption{The architecture of STAD aimed at generating SR EEG from LR EEG.}
    \label{fig:model}
\end{figure*}

In this paper, a novel EEG SR method, called spatio-temporal adaptive diffusion models (STAD), is proposed to achieve spatial SR reconstruction from LR EEG to HR EEG. Employing a diffusion learning strategy, This approach improves EEG spatial resolution by learning the latent mapping relationship between LR EEG and HR EEG. The primary contributions of this work are summarized as:

\begin{enumerate}
    \item A spatio-temporal adaptive framework based on diffusion models is proposed for EEG SR reconstruction. According to our knowledge, this marks the first time diffusion models are employed to achieve spatial SR reconstruction from LR EEG to HR EEG. This provides an efficient and cost-effective method for obtaining HR EEG, enhancing the portability and usability of HR EEG, and further advancing fields such as source localization and BCI technology.
    \item The spatio-temporal condition module (STC)  is designed to capture the spatio-temporal features of LR EEG. It further uses the generated spatio-temporal conditions to constrain the reverse denoising process, guiding the model to generate SR EEG that conforms to the characteristics of the subjects.
    \item The multi-scale Transformer denoising module (MTD)  is constructed to map LR EEG to HR EEG. It extracts temporal features at various scales and uses cross-attention-based diffusion Transformer blocks to adaptively modulate the denoising process based on spatio-temporal conditions, addressing the channel-level disparity between LR EEG and HR EEG.

\end{enumerate}

\section{Related Works}

In recent years, with the wide application of EEG in neuroscience and clinical diagnosis and treatment, how to improve the spatial resolution of EEG data has become a key problem to be solved. People try to adopt computer technology to enhance the spatial resolution of EEG, including channel interpolation methods based on mathematical models and spatial SR reconstruction methods utilizing deep learning.

\subsection{Channel Interpolation Methods}

The channel interpolation methods mainly utilize the spatial relationship between adjacent electrodes and signal characteristics for the reconstruction of missing electrode data to improve the spatial resolution. To estimate the values of missing channels based on the spatial relationships among electrode locations, Courellis \etal \cite{Courellis_inter} proposed an ellipsoidal geodesic length-based method to reconstruct missing or poor quality EEG signals by synthesizing the signals from neighboring electrodes. Petrichella \etal \cite{petrichell_inter} nvestigated two different interpolation methods to correctly reconstruct artifact-affected channel data and used to map global brain responses after stimulation of specific brain regions of interest. Nouira \etal \cite{nouira_inter} proposed a multiple quadratic interpolation technique for estimating scalp potential activity in 3-D EEG mapping. In this approach, by replacing the Euclidean distance between two electrodes with the arc length, multiple quadratic interpolation has been shown to be effective in reconstructing SR EEGs for subjects exhibiting various behavioral states.

\subsection{EEG SR reconstruction utilizing deep learning methods}

The deep learning-based spatial super-resolution reconstruction methods aim to learn the mapping relationship between LR EEG and HR EEG using deep learning models.
For example, Han \etal \cite{han_cnn} employed a deep CNN (DCNN) to enhance the spatial resolution of EEG data and showed that the SR method not only enhances the spatial resolution of EEG, but also improves the signal quality.
Tang \etal \cite{tang_gcn} proposed a deep EEG SR framework termed Deep-EEGSR, which combines a compact convolutional network and a fully-connected filter generation network (FGN). It leverages graph convolutions to adapt to the structural connectivity among EEG channels and employs sample-specific dynamic convolutions to adjust filter parameters to match individual functional connectivity patterns.
Saba-Sadiya \etal \cite{saba_ae} proposed a deep encoder-decoder network-based approach to automatically interpolate data loss caused by electrode “burst” artifacts in EEG, and used migration learning to enhance the robustness of the model.
El-Fiqi \etal \cite{el_ae} developed a weighted gating layer autoencoder for reconstructing lost EEG data. This new variant of the autoencoder focuses on learning the relationships between the variables, which can reveal the interactions between different channels.
Corley \etal \cite{corley_gan} further explored the application of GANs for EEG spatial super-resolution, aiming to reduce the cost of EEG devices by improving their spatial resolution.

\subsection{Diffusion Models}

In recent years, diffusion models have demonstrated remarkable generative capabilities across various domains, including image, audio, text, and medical data generation. Ho \etal \cite{ddpm} proposed the Denoising Diffusion Probabilistic Model (DDPM), which defines the fundamental components of diffusion models: the forward process and the reverse process, and demonstrated its superiority in image generation. Song \etal \cite{DDIM} introduced the Denoising Diffusion Implicit Model (DDIM), replacing the Markov forward process with a non-Markovian process, thereby improving the sampling speed of diffusion models without compromising the generative performance. Rombach \etal \cite{rombach2022LDM} applied diffusion models in the latent space, termed Latent Diffusion Models (LDMs), and utilized cross-attention mechanisms to incorporate conditional information into the underlying UNet architecture, guiding the model's conditional generation. Gao \etal \cite{gao2023implicit} proposed an implicit diffusion model (IDM) for continuous image super-resolution, which dynamically adjusts the ratio of real information in the low-resolution features to the fine details generated by the diffusion process through a scale-adaptive conditioning mechanism. Peebles \etal \cite{peebles2023DIT} proposed a Transformer-based diffusion model framework (DiT), replacing the UNet backbone in LDMs with a Transformer architecture and employing adaptive layer normalization for learning. Ozbey \etal \cite{ozbey2023SynDiff} introduced an adversarial diffusion model-based method (SynDiff) for cross-modality translation from Magnetic Resonance Imaging (MRI) to Computed tomography (CT), designing a cycle-consistent architecture for bidirectional modal conversion. Wang \etal \cite{wang2024diffmdd} proposed a diffusion-based deep learning framework (DiffMDD) for Major Depressive Disorder (MDD) diagnosis using EEG, incorporating a forward diffusion noisy training module to extract noise-irrelevant features for improved robustness, and a reverse diffusion data augmentation module to increase data size and diversity for learning generalized features.

\section{Method}

\subsection{Preliminaries and Problem Statement }

We will provide a brief review of some fundamental concepts essential for understanding diffusion models \cite{ddpm} before introducing our architecture. Diffusion models presuppose a forward diffusion process that incrementally adds noise to input data $x_0$ as
$q(x_t | x_0) = \mathcal{N}(x_t \mid \sqrt{\bar{\alpha_t}}x_{0}, (1 - \bar{\alpha_t}) I)$, where $\bar{\alpha_t}$ signifies the variance schedule for the amount of Gaussian noise.

In the reverse denoising process, the denoising model needs to learn the distributions $p_{\theta}(x_{t-1} | x_t)$ to denoise the each state $x_t$ during training. Furthermore, the training of diffusion models is accomplished using a variational method-based training strategy \cite{mo2009variational}. Formally, the inference procedure can be characterized as a reverse Markov process that starts with Gaussian noise $x_T \in \mathcal{N}(0,I)$, and progresses towards the target data $x_0$ as:
\begin{equation}
\begin{aligned}
    & p_{\theta}(x_{t-1} | x_t) = \mathcal{N} (x_{t-1} \mid \mu_{\theta}(x_t,t),\sigma_{t}^2I),
    \label{eqn:inverse}
\end{aligned}
\end{equation}
With $p_{\theta}$ trained, new target data can be obtained by initializing the variable $x_T \sim \mathcal{N}(0, I)$ and then using the reparameterization trick to sample $x_{t-1} \sim p_{\theta}(x_{t-1}|x_t)$ step by step, where $T$ denotes the number of forward process steps.

In this work, given an LR-HR EEG pair denoted as $(x, y)$ with $x \in \mathbb{R}^{C_{l} \times N}$ being a degraded version of $y \in \mathbb{R}^{C_{h} \times N}$, where $C_{l}$ and $C_{h}$ denote the number of the LR and HR EEG channels, respectively. $N$ is the length of each EEG epoch. Therefore, the process of mapping the LR EEG to the corresponding SR EEG can be described as
\begin{equation}
    y \approx y_{sr} = \mathit{f} (x,\theta)
    \label{eqn:eegsr}
\end{equation}
where $y_{sr}$ represents synthetic SR EEG. $\mathit{f}$ stands for the mapping function, while $\theta$ represents the optimal parameters sought to reconstruct the HR EEG.

\subsection{Basic ideas}

Learning the mapping relationship from LR EEG to HR EEG is challenging. To ensure that the generated results conform to the subject's characteristics, the proposed STAD use the spatio-temporal features of LR EEG as conditions and employs a cross-attention-based conditional injection mechanism to guide the denoising model for progressive denoising restoration. This mechanism embeds the spatio-temporal information of LR EEG into the high-dimensional latent space, indirectly guiding the model to comprehend the potential correlation between LR EEG and HR EEG. This ensures that the synthetic SR EEG is highly consistent with HR EEG in both temporal and spatial domains, addressing the issue of the large channel-level disparity between LR EEG and HR EEG.

As Fig. \ref{fig:model} shows, the entire model framework consists of EEG pre-trained autoencoder, STC and MTD. During the training phase of STAD, the HR EEG ground truth is first input into the EEG pre-trained encoder to obtain the corresponding latent vectors $z_0$. The forward diffusion process progressively derives the state $z_t$ of each time step $t$. In the reverse denoising process, the STC first encodes the time series and channel spatial distribution matrix from LR EEG into the spatio-temporal representation conforming to the subject, and concatenates it with time embedding as the conditioning vector.
MTD takes the conditioning vectors and noise sampled from the standard normal distribution as input, implements the conditional guidance mechanism through the cross-attention layer, and predicts the noise at each step of the reverse process. The Mean Squared Error (MSE) loss is minimized between the predicted and actual noise at each step, which guides the STC module and MTD module in performing iterative parameter updates during the training phase. During the sampling phase of STAD, the deterministic MTD precisely generates SR EEG matching the input LR EEG by progressively sampling using the spatio-temporal conditional information of LR EEG and noise sampled from the Gaussian distribution.

\subsection{Architectures}

\subsubsection{Pre-trained EEG Autoencoders}

LDMs \cite{rombach2022LDM} have demonstrated that mapping input data into a latent space through a pre-trained encoder can significantly enhance the quality of high-resolution natural image generation and improve performance on other downstream tasks. Following the work of \cite{dreamdiffusion}, we employ a Masked Autoencoder (MAE) for asymmetric latent space representation of HR EEG. Given that the input EEG consists of multi-channel time series signals, we utilize two-dimensional EEG data as input, where each row represents the time series data of a single channel, and each column represents the values of all channels at a specific time point. Initially, this module divides the EEG data of each channel into fixed-length windows and randomly masks them according to a specified percentage. The MAE then predicts the missing values using the surrounding contextual cues.
By reconstructing the masked signals, the pre-trained EEG encoder can learn the latent spatio-temporal representations of brain activity across different subjects. Detailed information on the architecture of the pre-trained EEG encoder can be found in \cite{dreamdiffusion}.

\subsubsection{Spatio-temporal Condition Module}

In EEG signals, each channel represents the signals collected by electrodes placed at specific locations on the brain, recording electrical activity from different brain regions \cite{T2}. The spatio-temporal condition module $\tau_\theta$ based on Transformer is constructed to capture the temporal correlations between time points and the spatial correlations between EEG channels. This module takes the time series from LR EEG and the spatial positions of the channels as inputs and outputs corresponding encoding vectors $c = \tau_\theta(x)$ as conditional information for the reverse denoising process, where $x \in \mathbb{R}^{C \times T}$ represents the LR EEG. By extracting the spatio-temporal features of LR EEG, this module provides richer and more precise contextual information for the denoising model.

As the yellow part of Fig. \ref{fig:model} shows, the STC is mainly composed of a spatial position embedding layer, a 1D convolution block, and a Transformer block. More specifically, the spatial position embedding layer the electrode spatial coordinates of the input LR EEG and effectively transmits the spatial structure information to the model. Then, we partition the input time series into different patches and process them with the 1D convolution block for the initial extraction of temporal features \cite{wu20183d}. Finally, the Transformer block concatenates the temporal features with position encoding and channel spatial embeddings to output the final spatio-temporal condition $c$. The 1D convolution block consists of a 1D convolution layer, batch normalization layer and a ReLU activation function. In the Transformer block, a multi-head self-attention (MSA) layer is followed by feed-forward networks, both of which are encapsulated with residual connections and layer normalization.

\subsubsection{Multi-scale Transformer Denoising Module}

EEG, limited by the number and placement of electrodes, struggles with spatial resolution. High-density EEG devices have shown promise in tasks like source localization and BCI. The multi-scale Transformer denoising module is designed to enhance the spatial resolution of existing EEG devices. Traditional diffusion models, which typically use UNet, are effective in image generation but limited in handling the long-sequence dependencies of time-series data like EEG. Inspired by \cite{peebles2023DIT}, the MTD employs a Transformer backbone to more effectively extract spatio-temporal features. Additionally, multi-scale 1D convolution blocks are introduced to capture multi-scale temporal features in the reverse denoising process. This structure allows the model to process signals across different frequency ranges simultaneously \cite{multiscale}.

As shown in the blue part of Fig. \ref{fig:model}, the MTD mainly consists of position encoding, multi-scale 1D convolution blocks and diffusion Transformer blocks. Each diffusion Transformer block includes a normalization layer, a MSA layer, a cross-attention layer and a feed-forward network. The MTD $\epsilon_\theta$ takes $z_t$, timestep $t$ and
spatio-temporal condition $c$ as inputs, outputting the predicted noise $\epsilon_\theta(z_t, t, c)$ at timestep $t$.
The multi-scale 1D convolution blocks extract temporal features at different scales from the latent vectors $z_t$ using 1D convolution layers with various kernel sizes and concatenate them along the feature dimension. This convolution path yields an output that can be represented as:

\begin{equation}
\begin{aligned}
    & \widetilde{h}_i\ =\ BN(Conv(z_t,k_i)), \\
    & \tilde{H}_t\ =Concat(\widetilde{h}_1,\ldots, \widetilde{h}_n)
\end{aligned}
\end{equation}
where Conv denotes performing a 1D convolution on each channel sequence and BN represents batch normalization. $\widetilde{h}_i$ denotes the output of different 1D convoluntion layer and $n$ denotes the number of 1D convoluntion layers. Subsequently, each row of $\widetilde{H}_t$ is treated as a token and input into the diffusion Transformer block. Through the MSA layer, information from different scales is integrated, and global spatio-temporal dependencies are captured. For the input $\widetilde{H}_t \in \mathbb{R}^{l \times d}$, where $d$ is the feature dimension and $l$ denotes the sequence length. The output $o_t$ of MSA layer can be expressed as

\begin{equation}
    o_t = \widetilde{H}_t + MSA(LN(\widetilde{H}_t))
\end{equation}
where LN denotes the layer normalization operation. Furthermore, condition $c$ is incorporated into Transformer through a cross-attention layer. The output of cross-attention layer can be expressed as

\begin{equation}
\begin{aligned}
    & Q_t = W_t^Q \cdot o_t, K_t=W_t^K \cdot c,V_t=W_t^V \cdot c \\
    & Attn = softmax(\frac{{LN(Q}_t)LN({K_t}^T)}{\sqrt{d_k}})V_t
\end{aligned}
\end{equation}
where $W_t^Q \in\mathbb{R}^{d \times d_o} ,W_t^K \in\mathbb{R}^{d\times d_c}$ and $W_t^V\in\mathbb{R}^{d\times d_c}$ are projection matrices with learnable parameters. $\sqrt{d_k}$ is the scaling factor. Finally, the fused feature representation is obtained by processing the attention results through a feed-forward network, and a standard linear decoder is used to output a predicted noise. This enables the reparameterization trick to obtain the denoised latent $z_{t-1}$ for the previous time step. This mechanism effectively integrates the temporal and spatial features of EEG, providing rich context for subsequent denoising tasks, thereby skillfully achieving the adaptive generation of SR EEG.

\subsection{Loss Function}

In the forward diffusion process, diffusion models progressively add Gaussian noise to the clean input $z_0$ to generate a series of noisy data ${z_t}^T_{t=1}$. In the reverse process, the traditional method is to train a denoising model $p_\theta(z_{t-1}|z_t)$ to predict the state $z_{t-1}$ of the previous time step. Diffusion models draw on the principles of variation method \cite{wang2008variational}, optimizing model parameters by maximizing a variational lower bound based on the evidence lower bound (ELBO):

\begin{equation}
\begin{aligned}
    \mathcal{L}_{ELBO}(\theta) & = \mathbb{E}_{q(z_{t-1}|z_0,z_t)}\left[\log \frac{p_\theta(z_t|z_{t-1})p(z_{t-1})}{q(z_{t-1}|z_0,z_t)}\right] \\
    & + \mathrm{KL}(q(z_{t-1}|z_0,z_t)||p(z_{t-1}))
\end{aligned}
\end{equation}
where KL denotes the Kullback-Leibler divergence. However, directly predicting $z_{t-1}$ may lead to a complex and unstable training process. To simplify the training process, the MTD $\epsilon_\theta$ is employed to predict the noise at each time step $t$, and the MSE loss is minimized to learn the model parameters between the predicted noise $\epsilon_\theta (z_t, t,  \tau_{\theta}(x))$ and actual noise $\epsilon$, where $x$ denotes the input LR EEG. The final loss function can be expressed as:
\begin{equation}
    \mathcal{L}_{DM}(\theta) = \mathbb{E}_{z,\epsilon \sim \mathcal{N}(0,1),t, x} \left[ \left\lVert \epsilon_t - \epsilon_\theta(z_t, t, \tau_{\theta}(x)) \right\rVert_2^2 \right]
\end{equation}
where $\epsilon_\theta$ is the proposed multi-scale Transformer denoising model and $ \tau_{\theta}$ represents the proposed STC module. Algorithm \ref{alg:train} presents the entire training procedure along with the final loss function.

\begin{algorithm}
\caption{Training Phase of STAD}
\label{alg:train}
\begin{mdframed}[backgroundcolor=bottomcolor,rightline=false,leftline=false,topline=false,bottomline=false,innerleftmargin=5pt,innerrightmargin=5pt,userdefinedwidth=0.92\linewidth,innerbottommargin=5pt,innertopmargin=5pt]

\BlankLine

\textbf{Input:}

$x$ \tcp*[r]{Preprocessed LR EEG data}

$y$ \tcp*[r]{Preprocessed HR EEG data}

$T$ \tcp*[r]{Number of diffusion steps}

$Epoch$ \tcp*[r]{maximal iterative number}

\BlankLine

Initialize parameters $\theta$ of the STAD

\For{$s \gets 1$ \KwTo $Epoch$}
{
    \textbf{Perform:} Obtain latent vectors via Encoder of MAE \;

    $z_0 = Encoder\left(y\right)$ \;

    \textbf{Perform:} Extract spatio-temporal features of LR EEG $x$ via STC $\tau_\theta$ \;

    $c = \tau_\theta\left(x\right)$ \;

    \BlankLine




    \textbf{Perform:} Forward Diffusion Process \;

    \For{$t \gets 1$ \KwTo $T$}
    {
        Sample $\epsilon_t \sim \mathcal{N}(0, I)$ \;
        $z_t \gets \sqrt{\bar{\alpha_t}} z_0 + \sqrt{1 - \bar{\alpha_t}} \epsilon_t$ \;
    }

    \BlankLine

    \textbf{Perform:} Reverse Denoising Process via MTD $\epsilon_\theta$ \;

    \For{$t \gets T$ \KwTo $1$}
    {
        $\hat{\epsilon_t} \gets \epsilon_\theta(z_t, t, c)$ \;
        $\hat{z}_{t-1} \gets \frac{1}{\sqrt{\bar{\alpha_t}}} \left( z_t - \frac{\beta_t}{\sqrt{1 - \bar{\alpha}_t}} \hat{\epsilon_t} \right)$ \;
    }

    \BlankLine

    \textbf{Perform:} Obtain SR EEG via Decoder of MAE \;

    $\hat{y} = Decoder\left(\hat{z}_0\right)$ \;

    \BlankLine

    \textbf{Loss Calculation:}

    $\mathcal{L}_{DM}(\theta) = {\left \| \epsilon_t - \epsilon_\theta(z_t, t, c) \right \|}^2_2$ \;

    \BlankLine

    \textbf{Update:}  Optimize $\theta$ via backpropagation \;
}

\textbf{Result:} The generated SR EEG $\hat{y}$.

\BlankLine

\end{mdframed}
\end{algorithm}

\section{Experiment}

\subsection{Dataset and Preprocessing}

In this work, the EEG dataset is collected from a publicly accessible dataset (Localize-MI \cite{dataset}), which consists of high-density EEG data collected from seven drug-resistant epilepsy patients (mean age=35.1, $sd$ age=5.4), comprising a total of 61 sessions (average session duration per subject=8.71, $sd$ session duration per subject=2.65). Additionally, the dataset contains the spatial locations of stimulus contacts in MRI space, Freesurfer surface space \cite{freesurfer}, and MNI152 space \cite{mni}for each subject, along with the digital coordinates of 256 scalp EEG electrodes.

\begin{table}[h]
    \centering
    \caption{The corresponding electrode systems for different scaling factors.}
    \begin{tabular}{ccc}
    \toprule[1.5pt]
    Scaling Factor     &  Channels & Montage \\
    \midrule[1pt]
    2     &  128 & EGI-128 montage\\
    4 & 64 & EGI-64 montage\\
    8 & 32 & EGI-32 montage\\
    16 & 16 & EGI-16 montage\\
    \bottomrule[1.5pt]
    \end{tabular}
    \label{tab:channel-level}
\end{table}

We employed the MNE software for preprocessing of the raw data. In this work, we applied high-pass filtering to the continuous EEG time series to remove low-frequency irrelevant information. To eliminate the large line noise caused by direct current, we used notch filtering at 50, 100, 150, and 200 Hz. Subsequently, we segmented the raw data into epochs of fixed duration (350 ms) based on the provided annotation events in the dataset. Moreover, the dataset was recorded using the EGI NA-400 amplifier (Electrical Geodesics) \cite{egi} at a sampling rate of 8000 Hz for 256-channel EEG data. Besides, we downsampled the original HR EEG data along the channel dimension to obtain LR counterparts at different spatial resolutions. As shown in Table \ref{tab:channel-level}, we set the SR scaling factors to 2, 4, 8, and 16. Specifically, each scaling factor corresponds to a specific number of channels from the EGI standard electrode system, thereby simulating the acquisition of EEG data with varying densities in real-world scenarios. Following the preprocessing steps, the paried dataset comprises a total of 23,930 samples. , with 80\%, or 19,144 samples, allocated to the training set for model learning, 10\%, or 2,393 samples, designated as the testing set to assess the model's generalization capabilities, and the remaining 10\%, also 2,393 samples, serving as the validation set for evaluation of model performance.

\subsection{Experiment Settings}

\subsubsection{Implement Details}

The proposed STAD are implemented using the PyTorch framework, and the corresponding experiments were conducted on an Nvidia RTX A800 GPU. With the Adam optimizer, training of the model was carried out using a learning rate of $2 \times 10^{-4}$, 300 epochs, and a batch size of 32. In the diffusion model framework, the time step of the forward process was set to 1000, during which the noise levels were modulated by sine or cosine schedules, and Gaussian noise was used for the noise type. In the multi-scale 1D convolution blocks, we employed 1D convolution layers with respective kernel sizes of 3, 5, 7, and 9. In the diffusion Transformer block, the number of attention heads was set to 16, and the hidden dimension was 64.

\subsubsection{Metrics}

To evaluate the quality of synthetic SR EEG, we employed four quantitative metrics: Pearson Correlation Coefficient (PCC),  Normalized Mean Squared Error (NMSE), Signal-to-Noise Ratio (SNR), and Mean Absolute Error (MAE). These metrics offer a thorough understanding of the similarity and reliability of the synthetic signals in comparison to the real signals.

The performance improvement of SR EEG in downstream tasks was quantitatively assessed using classification performance metrics including Accuracy (ACC), Precision, F1 Score, and Recall. Additionally, we used Localization Error \cite{hdeeg2} as a quantitative measure of performance improvement in source localization.

\subsection{Evaluation of EEG SR Reconstruction Performance}

In this section, the performance of the proposed STAD for EEG spatial super-resolution reconstruction is evaluated through various quantitative metrics and scalp topography maps.

\begin{figure}[ht]
\begin{minipage}[b]{1.0\linewidth}
    \begin{minipage}[b]{0.45\linewidth}
        \centering
        \centerline{\includegraphics[width=\linewidth]{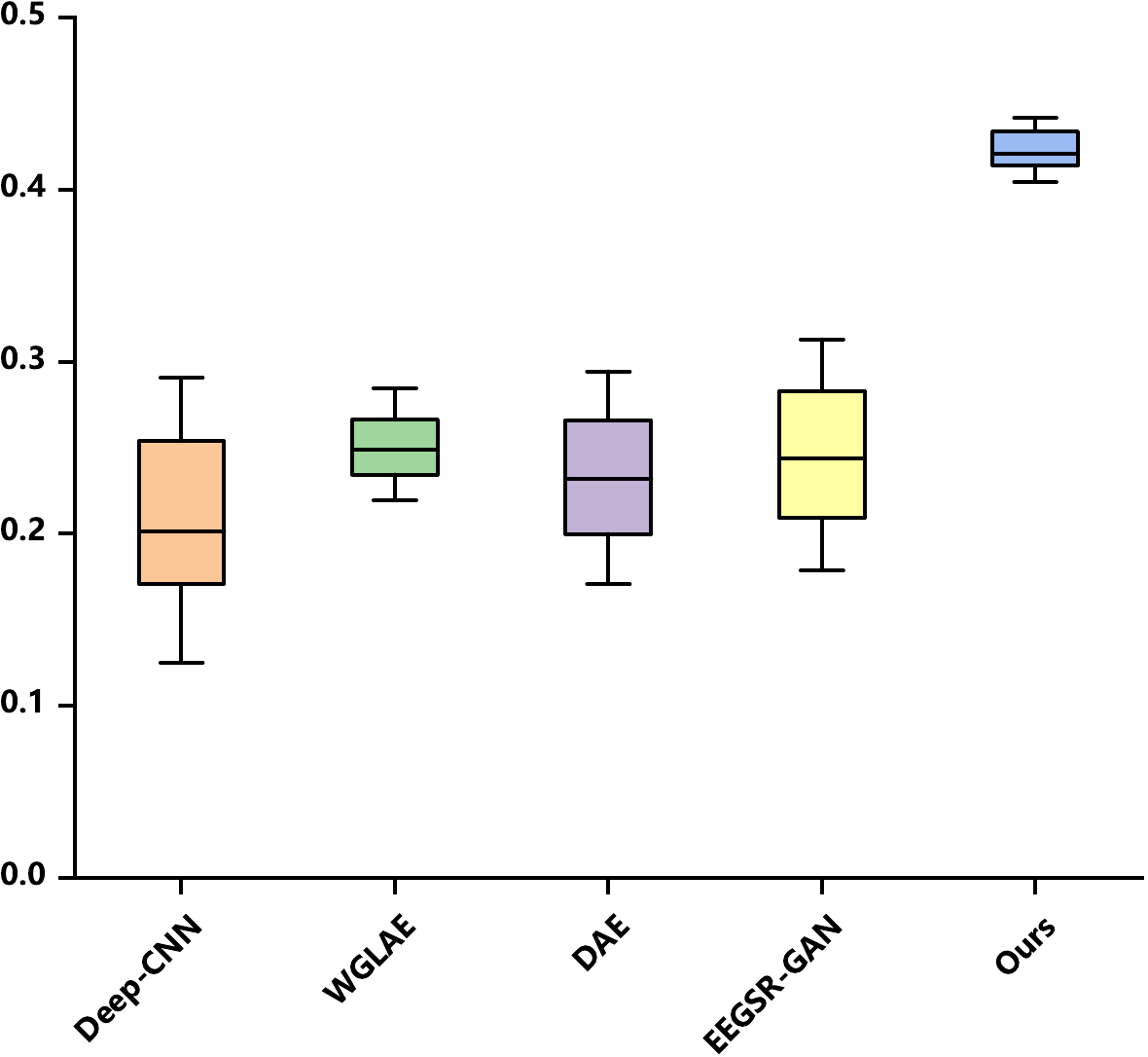}}
        \centerline{(a) PCC}\medskip
    \end{minipage}
    \hfill
    \begin{minipage}[b]{0.45\linewidth}
        \centering
        \centerline{\includegraphics[width=\linewidth]{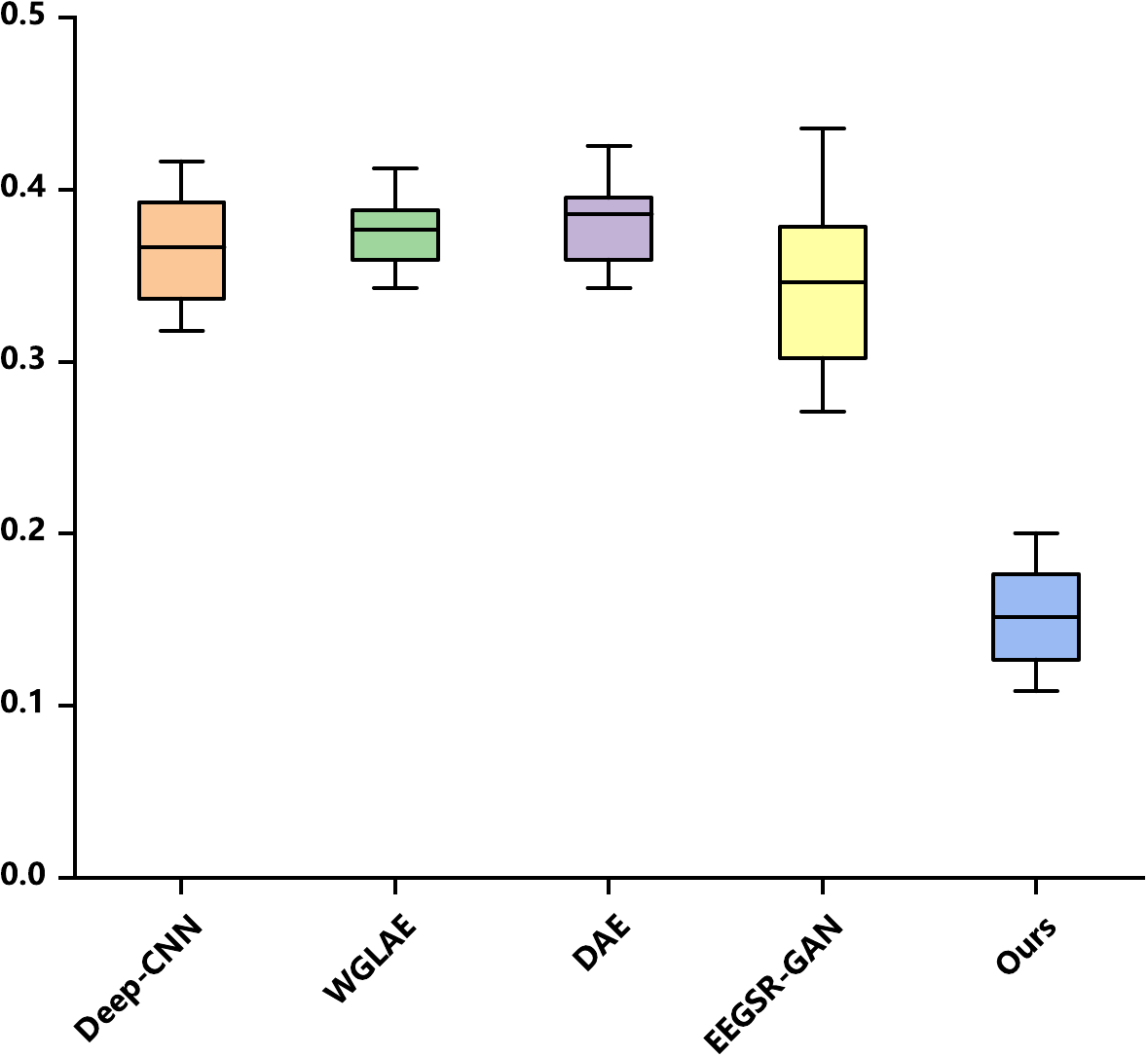}}
        \centerline{(b) MAE}\medskip
    \end{minipage}

    \begin{minipage}[b]{0.45\linewidth}
        \centering
        \centerline{\includegraphics[width=\linewidth]{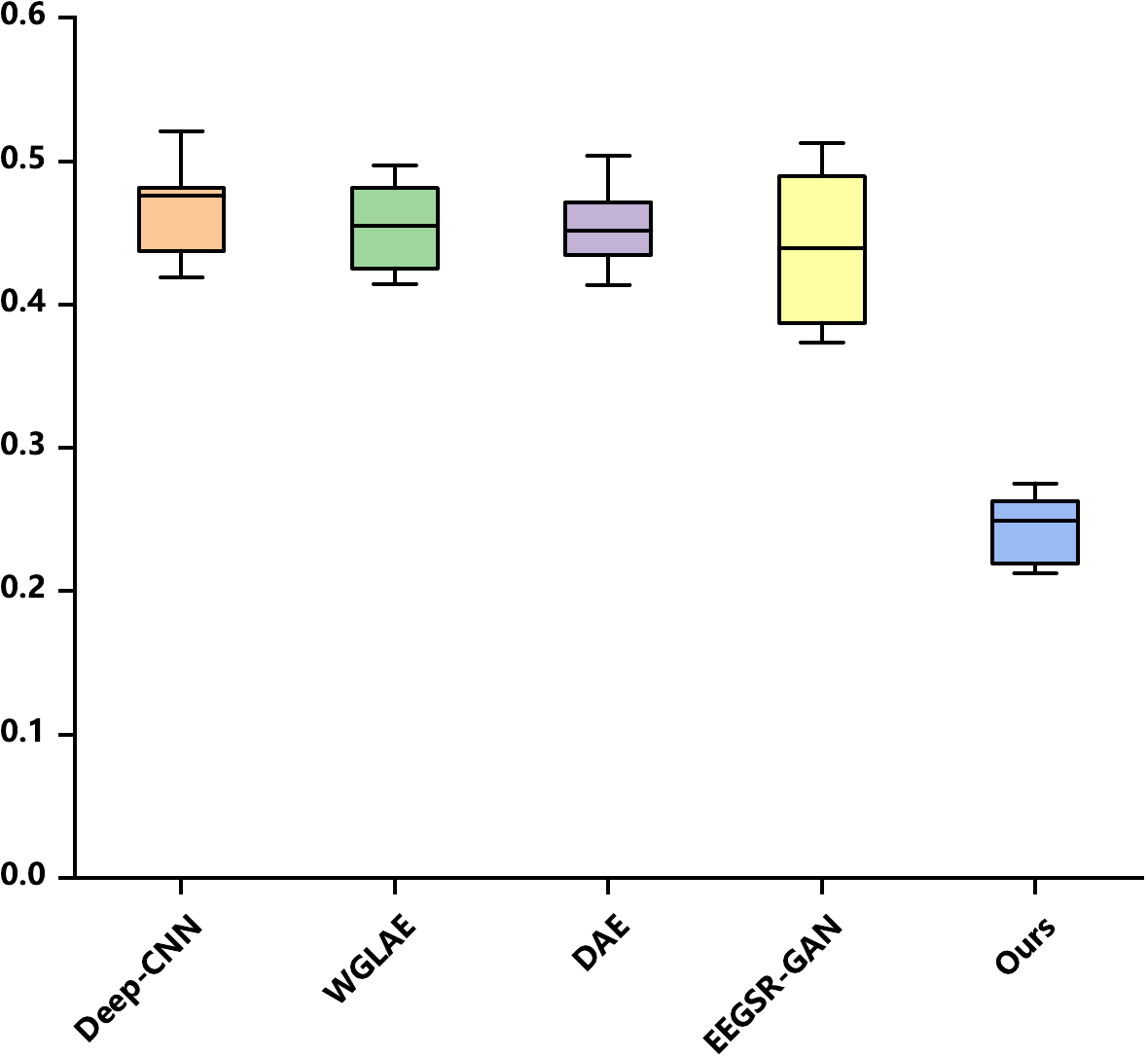}}
        \centerline{(c) NMSE}\medskip
    \end{minipage}
    \hfill
    \begin{minipage}[b]{0.45\linewidth}
        \centering
        \centerline{\includegraphics[width=\linewidth]{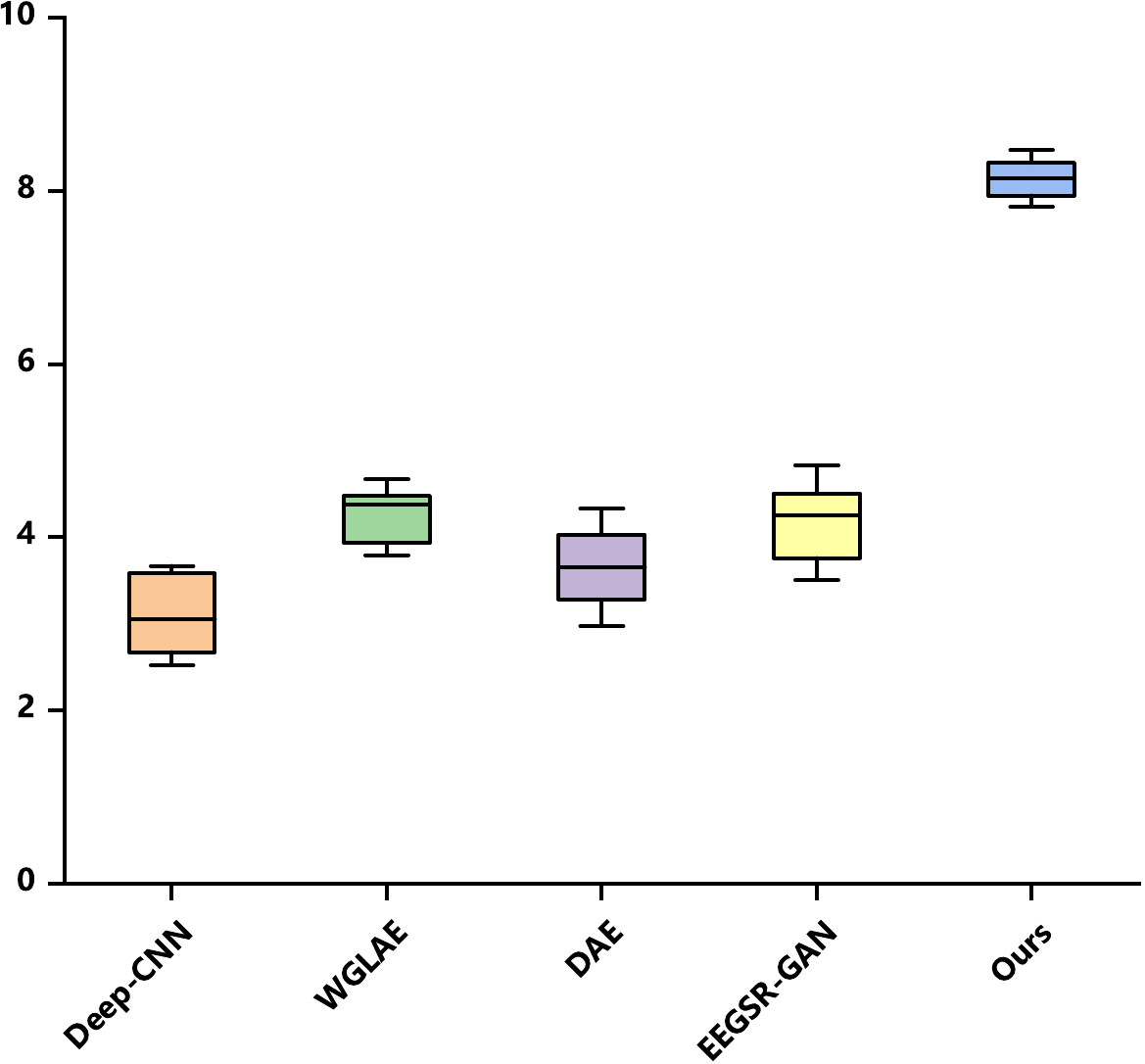}}
        \centerline{(d) SNR}\medskip
    \end{minipage}
\end{minipage}
\caption{Quantitative comparison results of STAD and existing EEG SR methods using four metrics. (a): The PCC performance across various methods. (b): The MAE performance across various methods. (c): The NMSE performance across various methods. (d): The SNR performance across various methods. The synthetic results of STAD demonstrate the highest quality across these four quantitative evaluation metrics.}
\label{fig:quantitative_comp}
\end{figure}

\subsubsection{Comparison with Other Existing EEG SR Methods}

the proposed STAD are compared with the existing EEG SR methods to demonstrate the effectiveness of the synthetic HR EEG (256 channels). We quantitatively analyze the reconstruction performance differences between the proposed STAD and several other methods: CNN-based methods (Deep-CNN \cite{han_cnn}), autoencoder-based methods [Weighted gate layer autoencoders (WGLAE) \cite{el_ae}, Deep autoencoder (DAE) \cite{saba_ae}], and  GANs-based method (EEGSR-GAN) \cite{corley_gan}.

Fig. \ref{fig:quantitative_comp} shows quantitative comparison results between different methods. As indicated, the proposed STAD exhibit superior performance, significantly surpassing other existing competitive methods. The highest PCC and SNR values indicate that the reconstructed results of STAD maintain high signal quality and correlation. Moreover, STAD significantly outperform other methods in terms of MAE and NMSE, indicating lower errors and higher accuracy in reconstructing SR EEG. Additionally, the gaps between STAD and other methods are substantial across all quantitative metrics. Through the experimental results, it can be concluded that compared to existing EEG SR methods, the proposed STAD effectively reconstruct high spatial resolution(256 channels) EEG, achieving significant spatial resolution enhancement of LR EEG.

\subsubsection{Comparison with Different Scaling Factors}

\begin{figure}[ht]
\begin{minipage}[b]{1.0\linewidth}
    \begin{minipage}[b]{0.45\linewidth}
        \centering
        \centerline{\includegraphics[width=\linewidth]{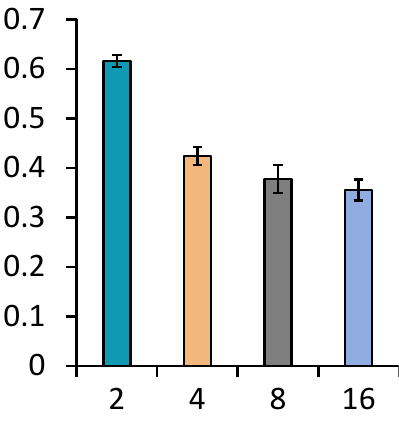}}
        \centerline{(a) PCC}\medskip
    \end{minipage}
    \hfill
    \begin{minipage}[b]{0.45\linewidth}
        \centering
        \centerline{\includegraphics[width=\linewidth]{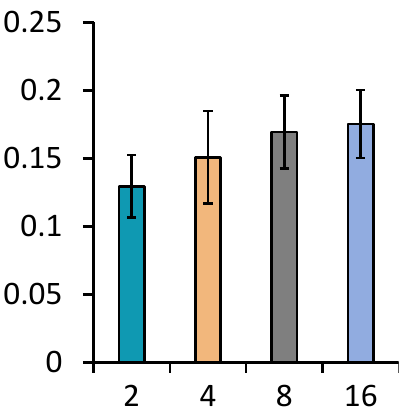}}
        \centerline{(b) MAE}\medskip
    \end{minipage}

    \begin{minipage}[b]{0.45\linewidth}
        \centering
        \centerline{\includegraphics[width=\linewidth]{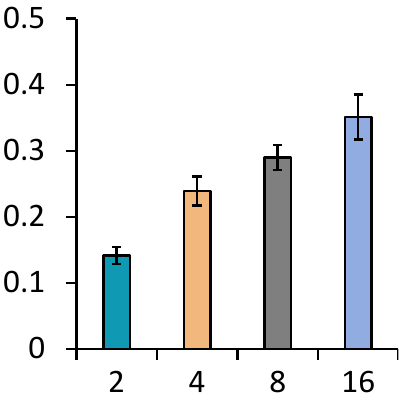}}
        \centerline{(c) NMSE}\medskip
    \end{minipage}
    \hfill
    \begin{minipage}[b]{0.45\linewidth}
        \centering
        \centerline{\includegraphics[width=\linewidth]{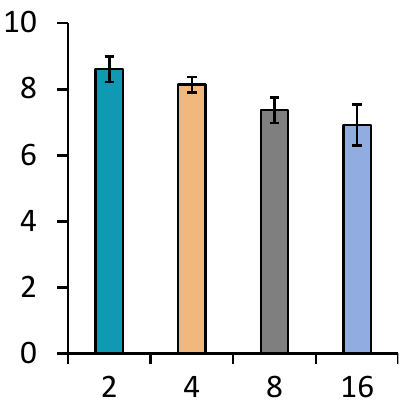}}
        \centerline{(d) SNR}\medskip
    \end{minipage}
\end{minipage}
\caption{Quantitative comparison between different scaling factors, including 2, 4, 8, 16. (a): The PCC performance across different scaling factors. (b): The MAE performance across  different scaling factors. (c): The NMSE performance across different scaling factors. (d): The SNR performance across different scaling factors.}
\label{fig:quantitative_scale}
\end{figure}

To evaluate the performance of the proposed STAD in generating synthetic SR EEG from LR EEG under different scaling factors, we employ four metrics for quantitative analysis of the model's generative performance. Fig. \ref{fig:quantitative_scale} shows quantitative results for different scaling factors. As the scaling factor increases, PCC and SNR values decrease, indicating a weakening correlation between synthetic SR EEG and ground truth. Meanwhile, MAE and NMSE values increase, implying a growing difference. The performance degradation of the model at high scaling factors can be attributed to two main reasons: Firstly, when larger scaling factors are used, the difference in the number of channels between LR EEG and HR EEG becomes more pronounced. Secondly, with fewer channels in the LR EEG input, the signal lacks sufficient temporal and spatial details, which makes it challenging for the model to recover the correct spatial structure and dynamic patterns during the reconstruction process.

However, even at larger scaling factors, the performance of STAD remains comparable to that at smaller scaling factors. For example, with a scaling factor of 16, the PCC value is only 3.02\% lower than at a scaling factor of 8, the MAE value differs by 1.68\%, and the SNR value by 3.19\%. This indicates that STAD can still adequately reconstruct SR EEG signals despite limited channel information. Overall, STAD effectively enhance LR EEG spatial resolution across different scaling factors, achieving better reconstruction at smaller scales while maintaining satisfactory performance at larger scales, demonstrating strong adaptability and robustness in handling various channel levels.

\subsubsection{Qualitative Results of Synthetic EEG}

\begin{figure*}[ht]
    \centering
    \includegraphics[width=\linewidth]{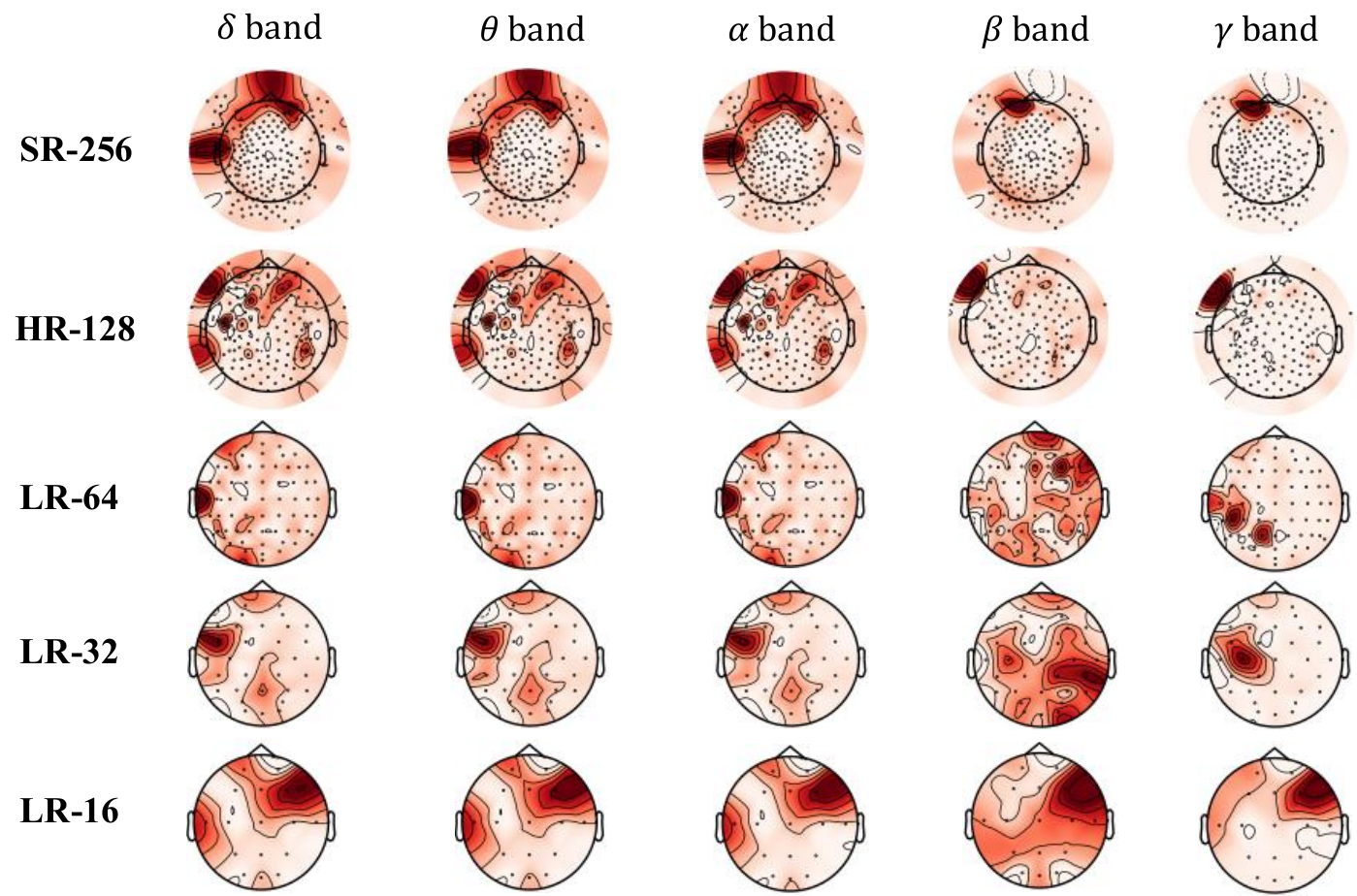}
    \caption{The comparison qualitative results of topomaps between synthetic SR EEG and LR EEG in different frequency bands.}
    \label{fig:topomap}
\end{figure*}

To observe the differences between synthetic SR EEG and LR EEG at different channel levels (16, 32, 64, and 128) more intuitively, we analyze the scalp topomaps of EEG in different frequency bands. Initially, the Multitaper method is employed to extract the power spectral density (PSD) of EEG with different spatial resolutions across five frequency bands, including $\gamma$ (30-40 Hz or higher), $\beta$ (13-30 Hz), $\alpha$ (8-13 Hz), $\theta$ (4-8 Hz), and $\delta$ (0.5-4 Hz) \cite{freqs}. Finally, we use the MNE tool to obtain scalp topographies in different frequency bands.
Fig. \ref{fig:topomap} compares scalp topographies across different frequency bands between EEG at various channel levels and synthetic 256-channel SR EEG. Each column represents a frequency band, and each row represents a type of EEG. As spatial resolution increases, the active regions on the scalp topographies become smaller and clearer. For instance, the 16-channel LR EEG shows large, blurry active regions in both low ($\delta$, $\theta$, $\alpha$) and high-frequency bands ($\beta$, $\gamma$), whereas the 256-channel SR EEG accurately localizes active sources in high-frequency bands. The synthesized SR EEG retains the frequency characteristics of the original signal across all bands, effectively reconstructing their spatial distribution. Furthermore, we observed that the synthetic SR EEG is able to retain the frequency band characteristics of the original EEG signal. This demonstrates the strong generalization of the model across different frequency bands and its ability to stably reconstruct EEG signals. In general, the proposed method generates SR EEG with richer spatial detail, more accurately describing and locating active regions on the scalp, while LR EEG results in coarse active regions, making it challenging to pinpoint active areas of the brain.

\subsection{Classification Evaluation on the Synthetic SR EEG}

Enhancing the spatial resolution of EEG significantly improves the performance of EEG-based diagnostic systems. For epilepsy patients, HR EEG aids doctors in identifying abnormal waveforms, such as spikes and sharp waves during seizures. To evaluate the effectiveness of synthetic SR EEG in detecting abnormalities, we designed a binary classification experiment (normal vs. abnormal) using SR EEG. In the Localize-MI dataset, EEG signals during stimulation are labeled abnormal, and those before stimulation are labeled normal. We trained the EEG-Net \cite{eegnet} separately on both LR EEG and the synthetic SR EEG and conducted corresponding classification tests. This allowed us to compare the performance of different spatial resolution LR EEG and synthetic SR EEG in the classification task.

\begin{figure*}[ht]
    \includegraphics[width=\linewidth]{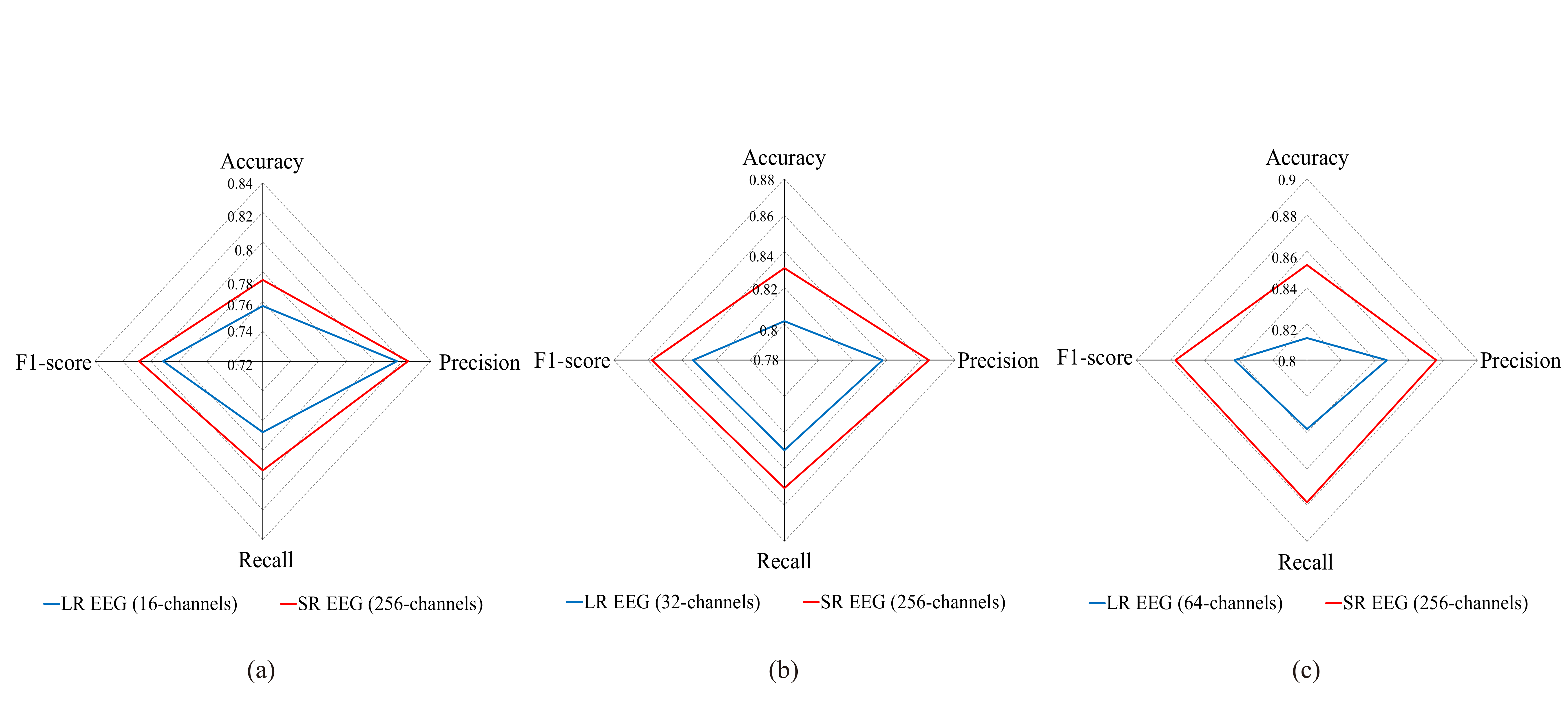}
    \caption{Evaluating classification performance across different spatial resolution of LR EEG and Synthetic SR EEG through radar charts. The assessment of classification performance relies on accuracy, precision, f1 score, and recall. (a): LR EEG(16 channels) versus SR EEG (256 channels). (b): LR EEG(32 channels) versus SR EEG (256 channels). (c): LR EEG(64 channels) versus SR EEG (256 channels).}
    \label{fig:radar}
\end{figure*}

Fig. \ref{fig:radar} shows the comparison results of four classification metrics between LR EEG and synthetic SR EEG. The SR EEG reconstructed by STAD outperforms LR EEG in all metrics, showing that STAD enhance classification performance of EEG-Net. As the channel count of LR EEG increases, the classification performance improvement by SR EEG also increases, indicating better detail reconstruction. This demonstrates that STAD can effectively enhance spatial resolution of LR EEG, improving downstream classification tasks. Furthermore, this provides a novel approach for other EEG signal enhancement and EEG-based clinical diagnosis research.

\begin{table}[ht]
\caption{The Clissfication Performance Comparison of Different Frequency Bands.}
\centering
\begin{tabular}{ccccc}
\toprule[1.5pt]
\multirow{2}{*}{\textbf{freq bands}} & \multirow{2}{*}{\textbf{EEG}} & \multicolumn{3}{c}{\textbf{scaling factor / Accuracy (AVG$\pm$STD \%)}} \\ \cmidrule(r){3-5}
                    &              & \textbf{4} & \textbf{8} & \textbf{16}  \\ \midrule[1pt]
\multirow{2}{*}{$\delta$ band}   & LR       & 69.81$\pm$0.02     & 67.21$\pm$0.00       & 63.87$\pm$0.12               \\
        & SR       & \textbf{74.42}$\pm$0.21     & \textbf{71.83}$\pm$0.33        & \textbf{69.42}$\pm$0.13            \\ \midrule[1pt]
\multirow{2}{*}{$\theta$ band}  & LR       & 71.12$\pm$0.04     & 69.78$\pm$0.12        & 64.15$\pm$0.00            \\
        & SR       & \textbf{75.36}$\pm$0.01     & \textbf{73.21}$\pm$0.32        & \textbf{72.13}$\pm$0.09             \\ \midrule[1pt]
\multirow{2}{*}{$\alpha$ band}  & LR       & 77.43$\pm$0.01     & 74.37$\pm$0.21        & 65.22$\pm$0.39             \\
        & SR       & \textbf{81.25}$\pm$0.04     & \textbf{79.54}$\pm$0.19        & \textbf{71.31}$\pm$0.04             \\ \midrule[1pt]
\multirow{2}{*}{$\beta$ band}  & LR       & 78.23$\pm$0.11     & 77.89$\pm$0.02        & 73.24$\pm$0.23           \\
        & SR       & \textbf{84.34}$\pm$0.16     & \textbf{81.47}$\pm$0.10        & \textbf{78.63}$\pm$0.12             \\ \midrule[1pt]
\multirow{2}{*}{$\gamma$ band} & LR       & 76.58$\pm$0.04     & 73.29$\pm$0.18        & 70.04$\pm$0.27            \\
        & SR       &\textbf{80.15}$\pm$0.12     & \textbf{78.19}$\pm$0.07        & \textbf{76.14}$\pm$0.17             \\ \midrule[1pt]
\multirow{2}{*}{all}
        & LR       & 83.92$\pm$0.24     & 81.59$\pm$0.24        & 77.85$\pm$0.15             \\
        & SR       & \textbf{87.61}$\pm$0.18   & \textbf{83.51}$\pm$0.28   & \textbf{79.52}$\pm$0.06             \\
\bottomrule[1.5pt]
\end{tabular}

\label{tab:classify}
\end{table}

We also analyzed classification performance from the frequency domain perspective, calculating the PSD of LR and SR EEG in five frequency bands. Each sample corresponds to $c \times5$ features (where c denotes the number of channels), and classification is performed based on these features. Table \ref{tab:classify} shows classification results for LR and SR EEG across different frequency bands and scaling factors. SR EEG outperforms LR EEG in all frequency bands. As the scaling factor increases, classification accuracy decreases for both but remains higher for SR EEG. Low-frequency bands ($\delta$, $\theta$, $\alpha$) have worse performance than high-frequency bands ($\beta$, $\gamma$), but the average accuracy improvement in low-frequency bands (5.06\%) surpasses that in high-frequency bands (4.95\%).
This finding can be attributed to the fact that low-frequency bands provide more accessible yet less comprehensive information, while high-frequency bands, despite being richer in detail, are more challenging to harness effectively due to their complexity and susceptibility to noise. These results suggest that the proposed STAD framework is adept at preserving low-frequency information while also capturing high-frequency variations, thereby enhancing the classifier's ability to identify abnormal EEG features with greater accuracy.

\subsection{Source Localization Evaluation on the Synthetic EEG}

Source localization is a crucial task in neuroscience and clinical applications, as it reveals the active source locations within the brain \cite{source}. To determine the impact of the proposed STAD in enhancing EEG spatial resolution and its effectiveness in improving downstream task performance, we assess it using the source localization task, incorporating both quantitative and qualitative analyzes. Specifically, we first use the subject’s MRI data to construct an individualized forward model. Then, we employ the eLORETA \cite{eLORETA1} as inverse solver algorithm which infers the distribution of brain current sources from scalp-recorded EEG signals through precise regularization strategies \cite{eLORETA2}. Furthermore, we use localization error \cite{hdeeg2} as the evaluation metric, defining it as the Euclidean distance between the source localization results and the ground-truth positions. This metric is employed to compare the performance of LR EEG, SR EEG, and HR EEG. For qualitative analysis, we visualize the source localization results on brain maps to provide an intuitive display of the localization performance across different data sources.

\begin{figure}[ht]
    \centering
    \includegraphics[width=\linewidth]{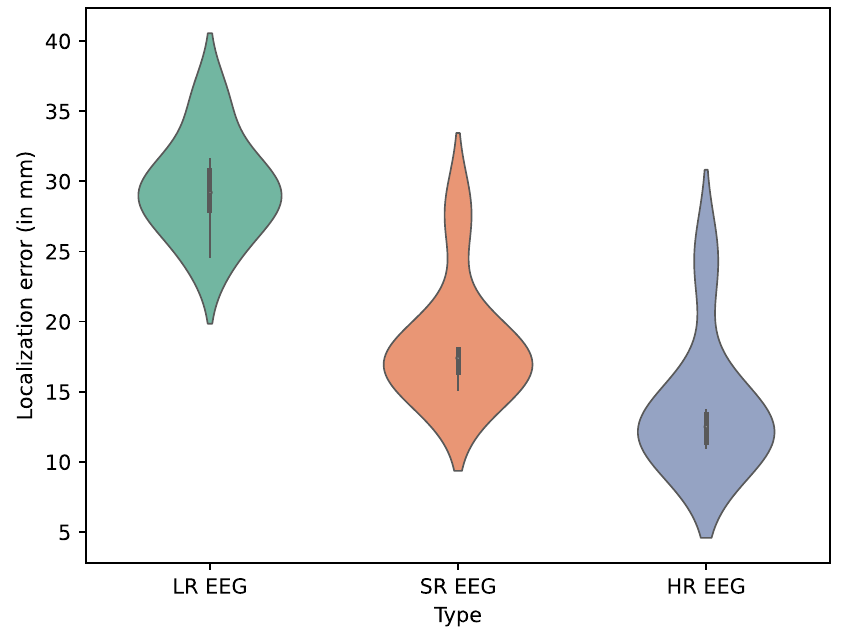}
    \caption{The localization errors across different EEG types for all Subjects.}
    \label{fig:locerror}
\end{figure}

Fig. \ref{fig:locerror} shows a comparison of localization errors for different EEG types among all subjects in the dataset. The localization error for LR EEG is widely dispersed with higher values, indicating lower precision and greater variability. In contrast, SR EEG shows significantly lower and more concentrated localization errors, suggesting higher precision and stability. Although the localization error of SR EEG is slightly higher than HR EEG, its accuracy is markedly improved, making it a viable substitute for HR EEG and reducing reliance on expensive high-density equipment. Therefore, the experimental results provide compelling evidence that the application of super-resolution techniques to LR EEG data substantially enhances the spatial resolution and source localization accuracy.

\begin{figure}[ht]
\begin{minipage}[b]{1.0\linewidth}
    \begin{minipage}[b]{0.45\linewidth}
        \centering
        \centerline{\includegraphics[width=\linewidth]{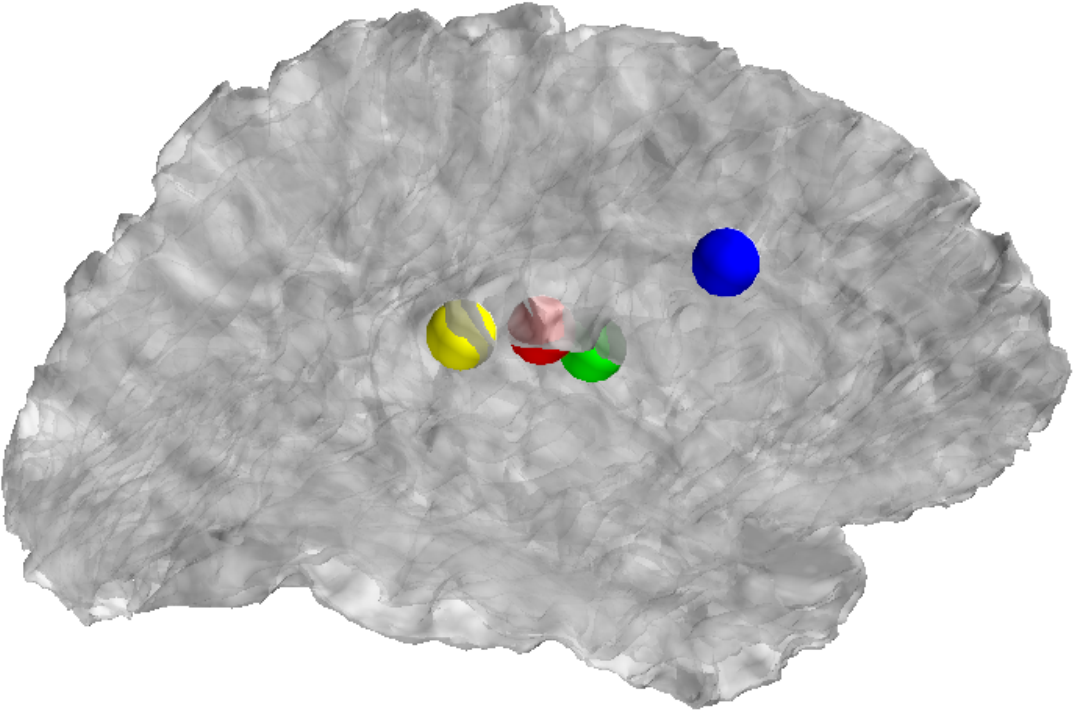}}
        \centerline{(a)}\medskip
    \end{minipage}
    \hfill
    \begin{minipage}[b]{0.45\linewidth}
        \centering
        \centerline{\includegraphics[width=\linewidth]{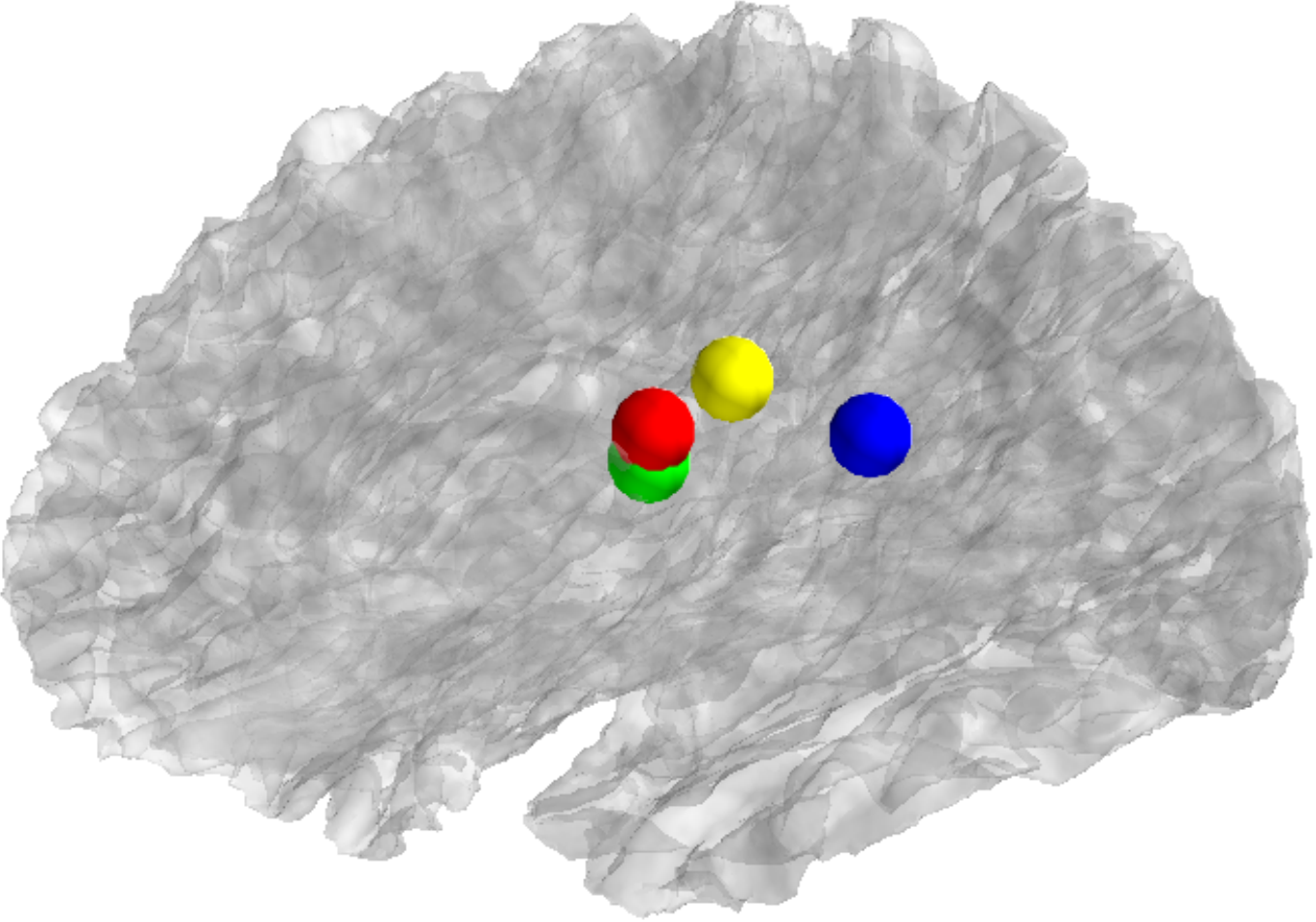}}
        \centerline{(b)}\medskip
    \end{minipage}

\end{minipage}
\caption{The visualization results of different EEG data types on source localization task.}
\label{fig:source}
\end{figure}

Fig. \ref{fig:source} shows the visualization results of different EEG data types on source localization tasks for some subjects. In these visualizations, green circles represent the stimulus source location ground-truth, red circles indicate the original HR EEG results, yellow circles denote the synthesized SR EEG results, and blue circles show the results from the LR EEG. The LR EEG results display considerable deviations from the ground truth, while the SR EEG results show minimal deviation, closely aligning with HR EEG performance. This demonstrates that the proposed method effectively enhances spatial resolution and signal quality, significantly improving source localization accuracy for LR EEG. Overall, By leveraging this method to augment the spatial resolution of LR EEG, researchers and clinicians can overcome the limitations imposed by low-density electrode configurations, facilitating more precise localization of neural sources and enabling a deeper understanding of brain dynamics.

\section{Discussion and Conclusion}

In this work, STAD are proposed to address the spatial resolution limitations of low-density EEG devices, particularly in clinical diagnostic applications, such as epilepsy focus localization. STAD are the first to employ a diffusion model to achieve spatial SR reconstruction from LR EEG to HR EEG. To ensure that the generated results align with subject-specific characteristics, a spatio-temporal condition module is designed to extract the spatio-temporal features of LR EEG, which are then used as conditional inputs to guide the reverse denoising process. Additionally, a multi-scale Transformer denoising module is constructed to leverage multi-scale convolution blocks and cross-attention-based diffusion Transformer blocks for conditional guidance, bridging the gap between SR EEG and HR EEG.
Qualitative and quantitative results indicate that STAD effectively enhance the spatial resolution of EEG. Furthermore, classification and source localization experiments demonstrate their ability to improve EEG performance in practical scenarios. In summary, By bridging the gap between low-cost, low-density EEG systems and the spatial resolution requirements of emerging neurotechnological applications, SR EEG paves the way for more accessible and precise neuroimaging, ultimately advancing our understanding of brain function and facilitating the development of innovative neurological interventions.

Experimental results show that while the proposed STAD outperform existing methods in reconstruction performance, larger scaling factors reveal a difference between the synthetic SR EEG and the ground truth, as illustrated in Fig. \ref{fig:quantitative_scale}. We identified several possible reasons for this: i) With larger scaling factors, the available spatial information for the model is constrained, making it challenging to accurately capture the spatial relationships between channels. ii) The inherently low SNR of EEG increases the difficulty of reconstruction for the model. iii) Significant individual differences between subjects make it challenging for the model to uniformly learn the mapping between HR EEG and LR EEG across different subjects. To address the issue of individual differences, personalized strategies can be designed, which is one of our future research priorities. Despite some limitations, the proposed method has demonstrated its effectiveness in downstream tasks and offers a new approach for enhancing the spatial resolution of EEG signals.

\begin{figure}[ht]
    \centering
    \includegraphics[width=\linewidth]{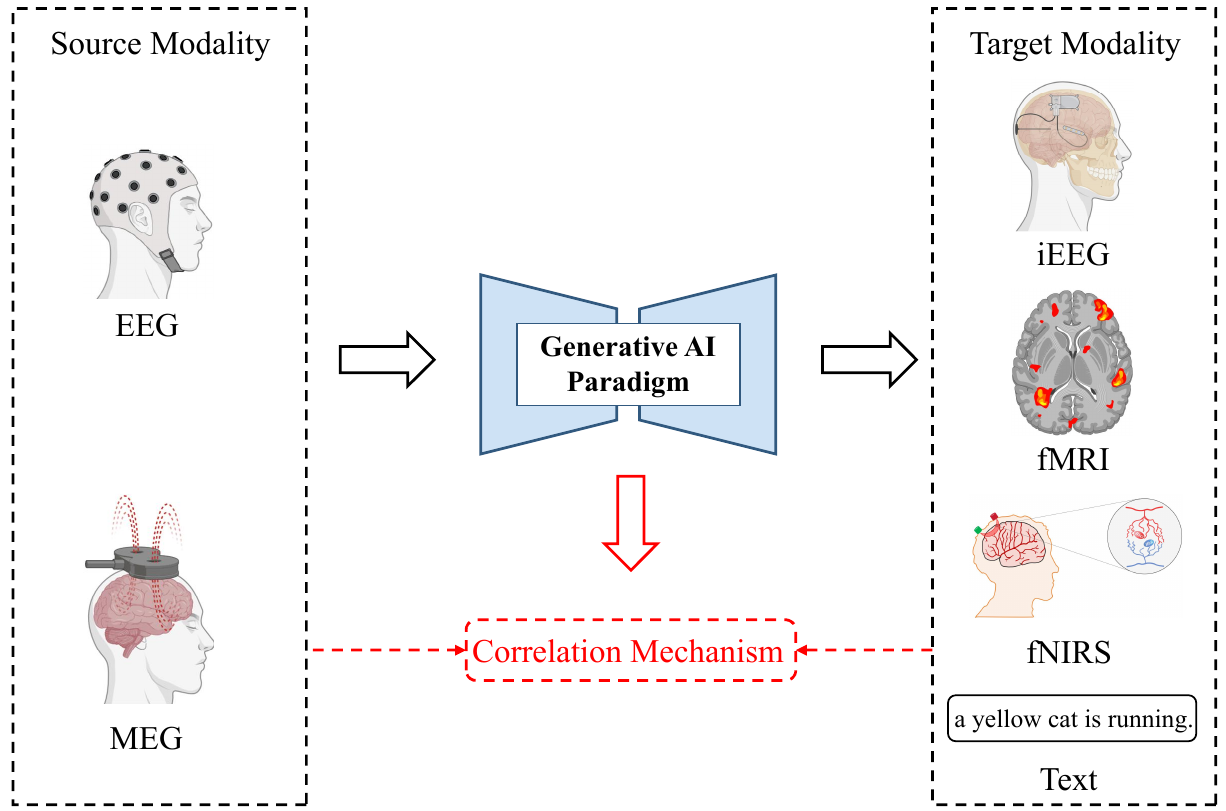}
    \caption{The framework for Constructing correlation mechanisms among different brain functional modalities using the generative AI paradigm.}
    \label{fig:discussion}
\end{figure}

In our future work, we aim to broaden the applicability of generative AI to enhance other brain function signals. As shown in Fig. \ref{fig:discussion}, we plan to employ this data-driven approach to address a broad range of neural signal tasks, including deep spatial super-resolution reconstruction of invasive EEG (iEEG), brain semantic decoding and the establishment of correlation mechanism among neural electrical signals, neural magnetic signals, and brain blood oxygenation signals, like fMRI and functional Near-Infrared Spectroscopy (fNIRS). Moreover, we will actively advance the integration of generative AI into real-world scenarios such as brain-computer interfaces, neurofeedback, and telemedicine. We anticipate that this diversified integration will not only propel research endeavors in neuroscience, cognitive science, and related fields but also enhance our understanding and diagnostic accuracy of neuropsychiatric disorders.

\section{Acknowledgment}

This work was supported by the National Key Research and Development Program of China under Grant 2023YFC2506902, National Natural Science Foundations of China under Grant 62172403, 12326614 and U2241210, and Strategic Priority Research Program of Chinese Academy of Sciences under Grant XDB38040200.

\bibliographystyle{ieeetr}
\bibliography{reference}

\end{document}